\documentclass[aps,pre,showpacs,showkeys,superscriptaddress,%
preprintnumbers,a4paper,%
twocolumn,twoside,floatfix%
]{revtex4}
\usepackage{amsfonts,amsmath}
\usepackage{bm}
\usepackage[dvips]{graphicx}
\usepackage{time}

\newcommand{\ie}{i.e.,\ }
\newcommand{\dd}{\ensuremath\mathrm d}
\newcommand{\R}{\ensuremath{\mathrm{Re}}}
\newcommand{\beq}{\begin{equation}}
\newcommand{\eeq}{\end{equation}}
\newcommand{\bea}{\begin{eqnarray}}
\newcommand{\eea}{\end{eqnarray}}

\begin{document}

\title{Lifetime statistics in transitional pipe flow}
\date{\today }
\author{Tobias M. Schneider}
\email{tobias.schneider@physik.uni-marburg.de} %
\affiliation{Fachbereich Physik, Philipps-Universit\"at Marburg, 
  Renthof 6, D-35032 Marburg, Germany}
\author{Bruno Eckhardt}
\email{bruno.eckhardt@physik.uni-marburg.de}
\affiliation{Fachbereich Physik, Philipps-Universit\"at Marburg, 
  Renthof 6, D-35032 Marburg, Germany}

\begin{abstract}
Several experimental and numerical studies have shown that turbulent
motions in circular pipe flow near transitional Reynolds numbers may not
persist forever, but may decay. We study the properties of these decaying
states within direct numerical simulations for Reynolds numbers up to
2200 and in pipes with lengths equal to 5, 9 and 15 times the diameter. 
We show that the choice of the ensemble of initial conditions 
affects the short time parts of lifetime distributions, but does not change 
the characteristic decay rate for long times. 
Comparing lifetimes for pipes of different length we notice a linear
increase in the characteristic lifetime with length, which reproduces
the experimental results when extrapolated to 30 diameters, 
the length of an equilibrium turbulent puff at these Reynolds numbers.
\end{abstract}

\pacs{47.27.Cn, 
47.52.+j, 
47.10.Fg, 
05.45.Jn, 
}
\maketitle 

\section{Introduction}
In several shear flows such as plane Poiseuille, plane Couette
\cite{Bottin1998, Barkley2005} and also pipe flow
\cite{Grossmann2000,Mullin2004,Kerswell2005, Eckhardt2007,Eckh08b} 
turbulent dynamics is observed for flow speeds where the laminar profile
is still stable against infinitesimal perturbations. In such 
a situation a finite amplitude perturbation is required
to drive the system from laminar to turbulent flow \cite{Grossmann2000,Boberg1988}, 
and one might expect that also for the converse process, returning from the
turbulent dynamics to the laminar one, a sufficiently 
large perturbation on top of the turbulent dynamics should be required. 
Several observations in direct numerical simulations and
experiments show, however, that turbulent motion returns to the
laminar flow suddenly and without any noticeable precursor or perturbation
\cite{Brosa1989,Darbyshire1995,Faisst2004,Hof2004}. From the point
of view of nonlinear dynamics such a behavior suggests that the 
turbulent state does not correspond to a closed off turbulent
attractor but rather to an open turbulent chaotic saddle 
\cite{Brosa1989,Schmiegel1997,Faisst2004}.  
One can then assign to each initial flow state a lifetime, i.e. 
the time it takes for this state to return to the laminar profile. 
The lifetime is a valuable observable that has previously
been used to extract information about states on the border
between laminar flow and turbulence \cite{Skufca2006,Schneider2007}. 
We will here use it to extract information about
the turbulent dynamics itself, thereby extending the work
reported in Ref. \cite{Faisst2004}.

Experiment and simulations show
that neighboring trajectories can have vastly different
lifetimes, so that the lifetime is rather unpredictable and depends sensitively on the
initial perturbation, see e.g.~\cite{Darbyshire1995,Faisst2004,Moehlis2004}.  
This strong sensitivity on initial conditions is consistent
with observations on other transiently chaotic systems and suggests that
rather than looking for the unpredictable behavior of individual 
trajectories, it is better to look for more reliable and stable properties
derived by averaging over ensembles of initial conditions. Prominent among 
such properties is the 
distribution of lifetimes, obtained from many runs with similar but not identical
initial conditions. The theoretical prediction for a hyperbolic saddle is that the
probability of decay is constant in time and independent of when the flow was 
started, giving for the distribution of lifetimes
an exponential, as in radioactive
decay \cite{Kadanoff1984,kant85,tel91}. Other functional forms
are possible as well (see e.g., \cite{Tel2008,Eckhardt2004}),
but for the most part observations in transitional shear flows
are compatible with an exponential 
\cite{Faisst2004,Hof2006,Mullin2006b,Peixinho2007,Bottin1998,Lagh07}.

An exponential distribution is characterized by a characteristic decay
rate or a \emph{characteristic lifetime} $\tau$ which is the time
interval over which the survival probability drops by $1/\mbox{e}$. 
How this lifetime varies with Reynolds number is currently under
debate \cite{Hof2006,Peixinho2006,Willis2007}. If $\tau$
diverges at a finite Reynolds number, there is a critical value
$\R_c$ above which turbulent flow does not relaminarize but persists
forever.  Such a divergence would imply that the system undergoes a
transition from a transient chaotic saddle to a permanently living
chaotic attractor in some form of `inverse boundary
crisis' \cite{Grebogi1982}. However, if $\tau$ does not
diverge, turbulence in a pipe remains transient for all $\R$. The
chaotic saddle does not close to form an attractor and the turbulent
`state' stays dynamically connected to the laminar profile even at
Reynolds numbers higher than the ones where `natural transitions' 
are reported to occur. 
This might open up new avenues for controlling turbulent motion.

The prediction of an exponential distribution of lifetimes is an 
asymptotic one, valid for long times. On short times the
distributions may follow a different functional form,  as evidenced by
the non-exponential parts in almost all distributions published so far. 
Moreover, the results may depend on additional parameters,
such as an aspect ratio or the length of the pipe. The dependence on these  
parameters has not been studied so far. It is our
purpose here to discuss some of these effects for transitional pipe flow.

We begin in section \ref{sec:survey} with a survey of previous
experimental and numerical results. Section \ref{sect:main} then is
devoted to an analysis of three effects: the dependence on the
ensemble of initial conditions in section \ref{sect:ensemble}, the
variation of the characteristic lifetime with $\R$ in \ref{sect:variation}
and the variation with the length of the pipe in section
\ref{sect:extensive}.  We conclude with a summary and outlook in
section \ref{sect:end}.

\section{Survey of results}
\label{sec:survey}
As usual, the mean downstream velocity $\langle{u}\rangle$, 
the diameter $D$ of the pipe and the viscosity $\nu$ of the fluid
can be combined into the dimensionless Reynolds number,
\beq 
\R=\frac{\langle{u}\rangle D}{\nu}\,.
\eeq 
The pipe diameter $D$ and the velocity $\langle{u}\rangle$ 
then define a unit of time $D/\langle{u}\rangle$. Since the
flow moves downstream with the mean velocity $\langle u\rangle$,
time can be translated into distance traveled, so that
the distance in units of the diameter equals the time $t$ 
in units of $D/\langle{u}\rangle$.
Because of this relation between length and time, it is crucial
to work with very long pipes so that the observation times become
as large as possible. 

When the flow becomes turbulent, the friction factor increases. Therefore,
either the forcing (pressure drop) has to increase so as to maintain the 
mean flow speed, or the mean flow speed will decrease, perhaps reducing
the Reynolds number so much that the flow relaminarizes \cite{Meseguer2007}. 
Thus, many modern
experiments work with a constant flow rate \cite{Darbyshire1995}, or with 
very long pipes \cite{Hof2006}, in which the change in Reynolds number becomes 
negligible as long as the turbulence remains confined to a small section of the
pipe: In the range of Reynolds numbers studied here, the turbulence is
localized in a region of about 30D length \cite{Wygnanski1973}.
To measure lifetime statistics one can either follow a puff on its
journey down the pipe and determine the downstream position where it 
decays; or one can choose a fixed downstream position, which
corresponds to a fixed lifetime, and measures the probability that
puffs survive the journey down the pipe up to this chosen point.  

The first approach was chosen by Mullin in a recent series of
experiments \cite{Mullin2006a,Mullin2006b,Peixinho2006,Peixinho2007} 
inspired by the numerical studies in \cite{Faisst2004}.  
In a first group of experiments \cite{Mullin2006a} the decay
of the perturbation could be detected with a camera that traveled
with the perturbation downstream. The length of the pipe allowed for a 
maximal observation time of 500 units. 
The flow was perturbed by injecting six 
jets of different amplitudes, and $40$ to $100$ 
independent repetitions were taken for each Reynolds number.
The asymptotic regime of the distributions of
lifetimes was found to follow a law 
\beq 
P(t) \sim
\exp\left(\frac{t-t_0}{\tau(\R)}\right)\;, 
\eeq 
with a characteristic
lifetime $\tau(\R)$ depending on $\R$ and an initial offset $t_0
\approx 100 \ldots 150$ before which no decay was observed. The strong
increase of the characteristic lifetime with $\R$ lead to the
conclusion that it diverges at a finite critical Reynolds number.
For the critical Reynolds number they give in \cite{Mullin2006a} the
values $\R_c = 1710 \pm 10$ and $\R_c =1830\pm 10$, and in
\cite{Mullin2006b} the values $\R_c = 1695 \pm 20$ and $\R_c = 1820\pm 20$,
for two different kinds of perturbations, described as `strong' and `weak' 
types of perturbation, respectively.
In an effort to address the dependence on the type of initial perturbations,
they performed a second experiment with a
slightly different perturbation protocol \cite{Peixinho2006}: 
In order to obtain more generic initial
conditions the system was started at a higher flow speed, a
perturbation that triggered turbulence was introduced and 
then the Reynolds number was reduced to the one for which lifetime statistics 
were collected. This gives another sample of initial conditions but limits 
the remaining observation time to less than 450. 
With such a perturbation the characteristic lifetimes were
compatible with 
\beq
\tau(\R) \propto (\R_c - \R)^{-1 \pm 0.02} \,,
\label{divergence}
\eeq 
but now with a different critical Reynolds number
of $\R_c = 1750\pm10$.

It is difficult to model the perturbations induced by jets in
numerical simulations (\cite{Meseguer2007}), but 
it is relatively straightforward though time-consuming to imitate
the second protocol of Mullin, where the initial conditions are taken
from a turbulent flow at higher Reynolds numbers. Willis and
Kerswell \cite{Willis2007} 
did just that for five different Reynolds numbers and concluded
that $Re_c=1870$, as suggested by some experiments. However, when
the analysis of their data points is corrected as suggested in \cite{arxiv},
the demonstration of a divergence is less convincing and the data
become compatible with the results of Hof et al \cite{Hof2006}. 

The experiments by Hof et al. \cite{Hof2006} just mentioned use a different approach. 
In a pressure
driven flow through a thin pipe of only $4$ mm diameter but $30$ m length they
realized dimensionless observation times of up to 7500 units. 
Since the flow could not be visualized, the time and position 
of decay could not be determined directly.
However, a laminar and a turbulent
patch in the flow can easily be distinguished once they leave the pipe, so that it
is relatively easy and straightforward to determine whether the flow has stayed
turbulent until it exits the pipe. Therefore, they could determine the probability
to be turbulent after a time period given by the distance between the perturbation and
the outlet, as a function of flow rate.
This gives $P(t,\R)$ as a function of $\R$ for $t$ fixed. Collecting
data for different $t$ then gives the parameters in the lifetime distribution
including the Reynolds number dependence. For short times, the data
are within the error bars of \cite{Peixinho2006}, but for longer times they
deviate from the divergent behavior implied by (\ref{divergence}). Instead, it
was found that the lifetimes are well represented by an exponential variation,
\beq
\tau^{-1}(\R) \propto \exp(a+b \R)
\label{exponential}
\eeq
with $a=55.3$ and $b=-0.032$.

\section{Lifetime distributions and their properties}
\label{sect:main}
In this section we study lifetime distributions in pipe flow within
direct numerical simulations. Since the calculations are extremely
time consuming, we will not aim to repeat the puff simulations of
\cite{Willis2007}, but rather focus on short, periodically continued
pipe sections, and then discuss how these results scale up to
turbulence in regions of the length of turbulent puffs.  In the next
subsection we first discuss features of individual turbulent
trajectories, before turning to the ensemble dependence of lifetime
distributions, the variations with Reynolds number and the length
dependence.

Individual trajectories were generated using the pseudospectral 
DNS code developed in \cite{Schneider2005} and already used in 
our previous studies \cite{Hof2006,Schneider2007,Schneider2007a,Eckhardt2007}.
Simulations of elongated puffs with the determination of their
travel velocity, envelope and internal dynamics are given in \cite{statphys}.
The code uses Fourier modes in downstream and azimuthal direction and
Chebyshev polynomials in the radial direction, and a projection method
to eliminate the pressure. The simulations on pipe segments 
presented in this section are carried out with $n$
Fourier modes in azimuthal and $m$ Fourier modes in downstream
direction, where ${|n|}/{N_{max}} + {|m|}/{Z_{max}} \le 1 \,$
with $N_{max}=16$
and $Z_{max}$ increasing from $14$ for the `short' pipe of length $5D$ to
$Z_{max}=25$ and $Z_{max}=45$ for the `medium' ($L=9D$) and `long'
($L=15D$) pipes, respectively. 
Consequently, we consider up to $33$ Fourier modes
in azimuthal direction. In downstream direction up to $31$
modes are considered for the short, $51$ for the medium and $91$ for
the longest pipe.
We use $49$ Chebyshev polynomials for the
expansion in the radial direction. 
This moderate resolution results from a compromise of accurate
representation of the dynamics and maximum simulation speed, required
for good statistics.

\subsection{Features of individual trajectories}

\begin{figure}
\begin{center}
\includegraphics[width=0.48\textwidth]{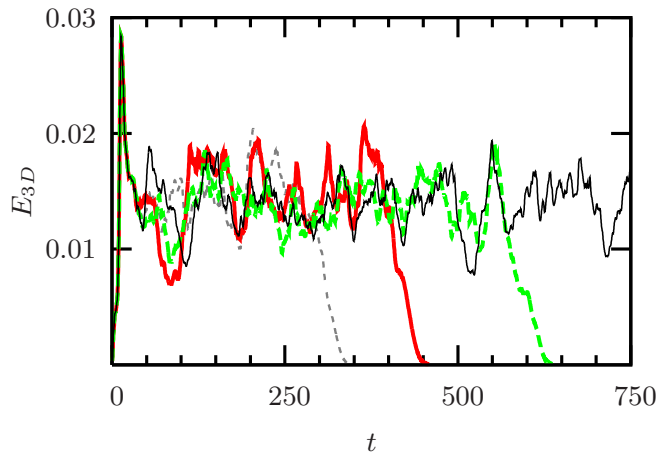} 
\end{center}
\caption[Energytraces at $\R=1900$, $L=15 D$ pipe]{(Color online) 
  Simulated evolution of four initial disturbances at $\R=1900$ in a
  periodic pipe of length $L=15D$. Plotted is the kinetic energy of
  the deviation from the laminar profile that is stored in the
  streamwise varying Fourier modes, eq.~\ref{energy_content}.  Initial
  conditions are constructed from a modulated Zikanov mode discussed
  in the main text. The four initial conditions differ by less than
  $0.5\%$ in energy content. The pertubations first grow in energy and
  show an overshoot before settling down to the turbulent state. They
  then suddenly decay without any prior indication and the energy of
  the perturbation decays monotonically.  The chosen criterion for
  decay is based on the energy threshold $E_{3D} < 5 \times10^{-5}$.
\label{fig:LT_energytraces}}
\end{figure}

Consider a perturbation of the laminar Hagen-Poiseuille flow applied
at time $t=0$.  The evolution of the initial condition $\vec{u}(t=0)$
can be followed in time until it decays or reaches the maximum
integration time in a simulation or leaves the pipe in an experimental
setup. Fig.~\ref{fig:LT_energytraces} shows the evolutions of four
different but similar initial conditions. As an indicator for
the turbulent intensity, we take the energy of the
three-dimensional structures,
\begin{equation}
  E_{3D}=\left({\sum_{m \ne 0} \int_{\text{Vol}} \vec{v}^2_{n,m} \dd
      V}\right)\left/ \left({\int_{\text{Vol}} 4(1-r^2)^2 \dd V}\right)
\right.\;,
\label{energy_content}
\end{equation}
  where $\vec{v}^2_{n,m}$ denotes the $(n,m)$-Fourier mode if the
  perturbation field $\vec{v}=\vec{u}-2(1-r^2)\vec{e}_z$ is decomposed
  into Fourier modes in azimuthal ($m$) and axial ($n$) direction. The
  energy content of the streamwise modulated Fourier modes is
  normalized by the kinetic energy of the laminar profile.

Since a flow field only
asymptotically reaches the laminar profile exactly, `decay' is defined
as reaching a situation where perturbations of the laminar profile are
so small that the further evolution follows an exponential drop off.
Any perturbation is therefore characterized by a lifetime that
slightly depends on the chosen criterion to detect being close to the
laminar profile.  Technically, one introduces a cut-off threshold
either on the kinetic energy stored in the deviation from the laminar
profile $\int ||\vec{u} - 2(1-r^2)\vec{e}_z||^2\dd V$, or on the
kinetic energy (\ref{energy_content}) 
stored in streamwise invariant Fourier modes only. The threshold
on these energies is chosen such that the further evolution can be 
described by the linearized equations, so that the system cannot
return to the turbulent dynamics. The lifetime then is defined as the
time it takes to reach this target region around the laminar profile.


%

\subsection{Ensemble dependence}
\label{sect:ensemble}

Now consider an ensemble of several different but similar
perturbations. The collection of individual lifetimes can be used to
estimate the probability $P(t)$ to still be turbulent at some time
$t$. 
A chaotic saddle should give rise to exponential asymptotic tails
of this distribution that are independent of the choice of initial
conditions but characteristic for the saddle.

\begin{figure}
\includegraphics[width = 0.18\textwidth]{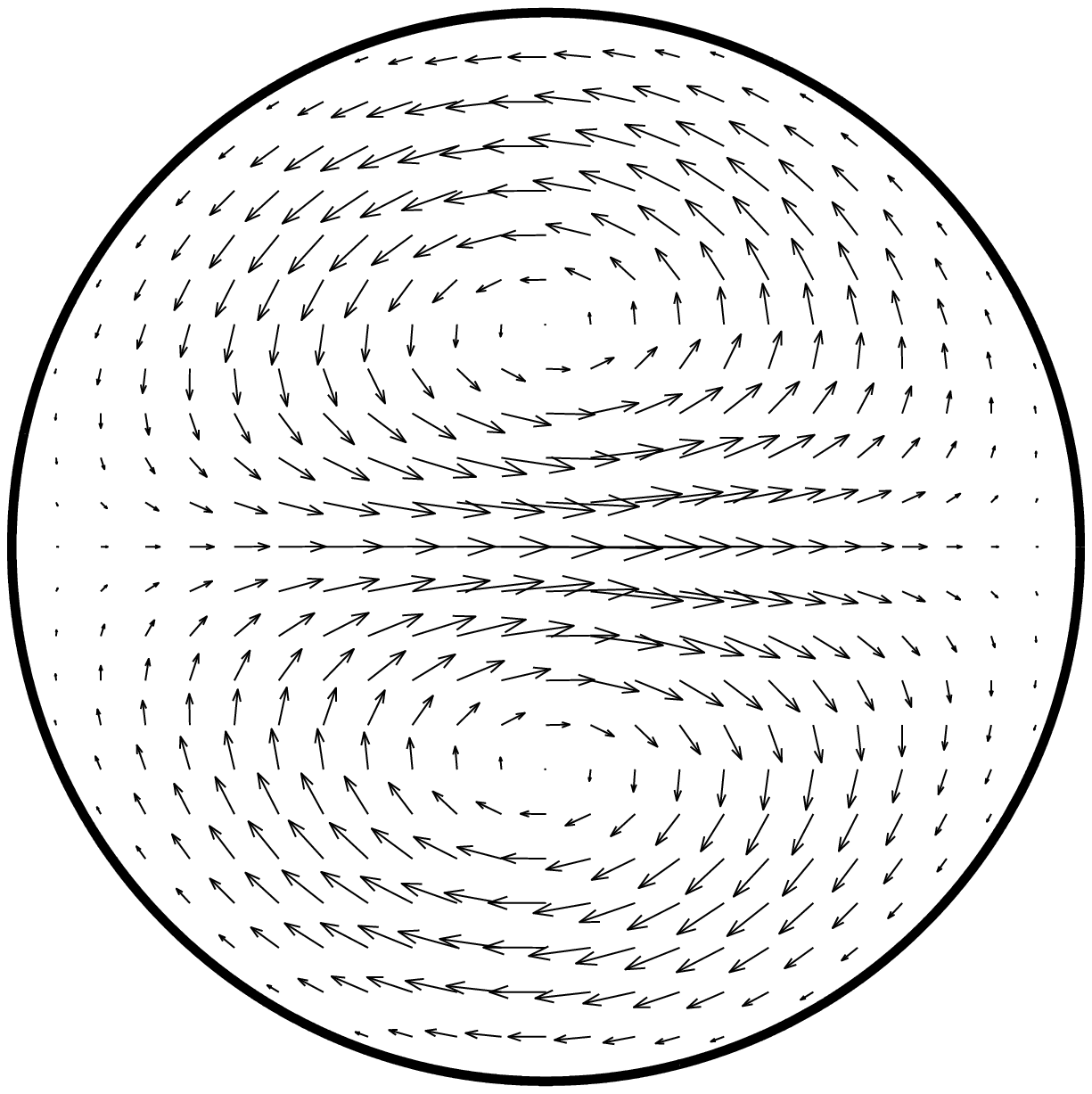} %
\includegraphics[width = 0.25\textwidth]{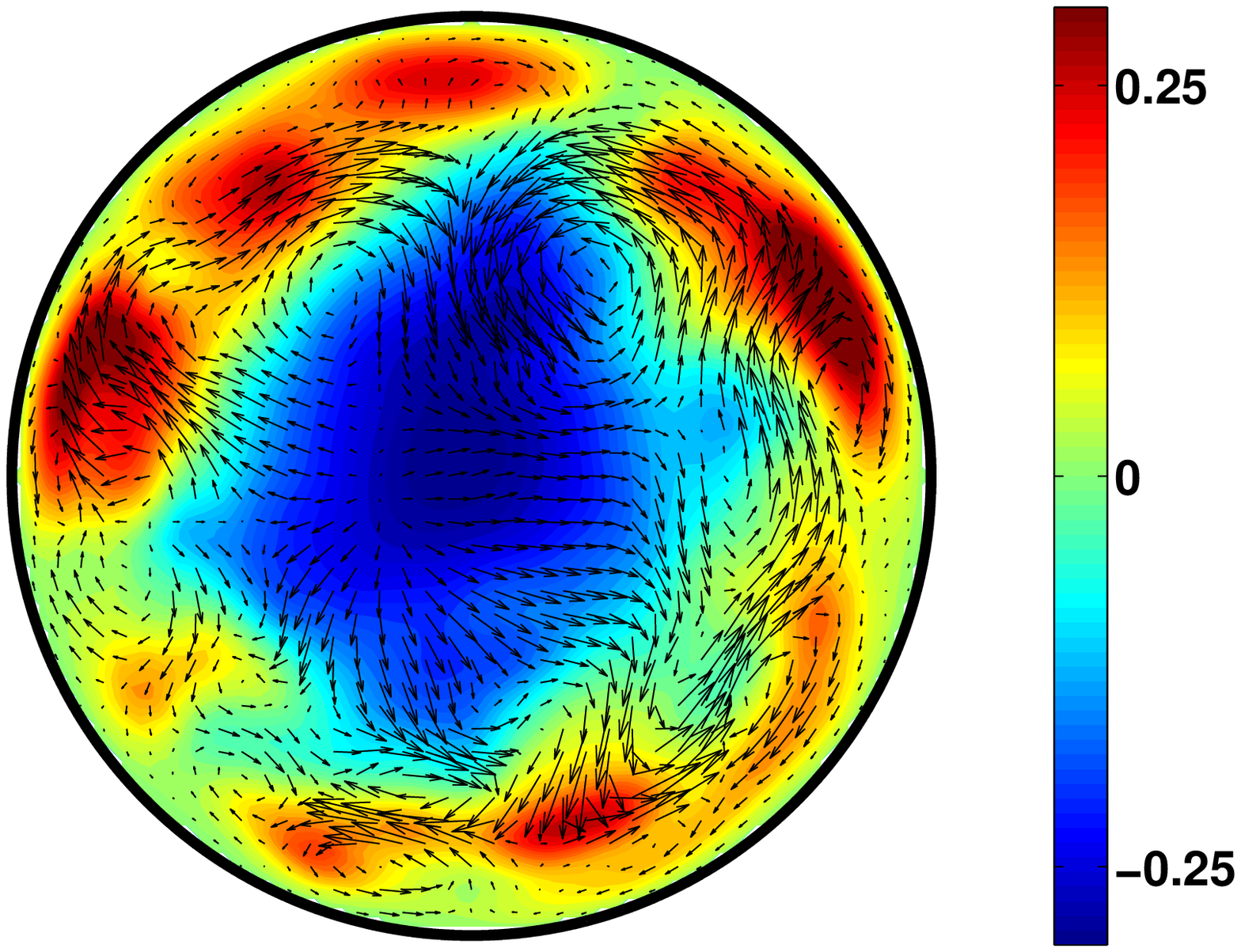} 
\caption[]{(Color online) 
  Spatial structure of the types of initial perturbations considered.
  Left panel: In plane velocity components of Zikanov's nearly optimal
  growing mode. Axial velocity components 
  vanish. The translational symmetry of Zikanov's mode is broken by
  applying a $z$-dependent twist. Right panel: Cross-section of a
  velocity field from a turbulent run at $\R=2150$. The vector plot
  indicates the in-plane motion. Color coding is used for the axial
  velocity relative to the laminar parabolic
  profile.\label{fig:LT_IC}}
\end{figure}

One type of perturbations we consider here is a pair of vortices as in
the optimally growing modes identified by Zikanov
\cite{Bergstroem1993,Schmid1994,Zikanov1996}.  In order to break
translational symmetry they are modulated in streamwise direction by
applying a $z$-dependent twist:
\beq 
\vec{u}_0(r, \varphi,z) =
\vec{u}_{\text{Zik}} \left(r\,, \varphi + \varphi_0 \sin\left(\frac{2
      \pi}{L} z \right),\,z\right)\;, 
\eeq 
where $\vec{u}_{\text{Zik}}$ is Zikanov's mode and $L$ is the length
of the computational domain used in our direct numerical simulation.
The spatial structure is presented in Fig.~\ref{fig:LT_IC}. Mimicking
experimental protocols, where the spatial structure of the
perturbation is prescribed by the setup, the ensemble of initial
conditions is constructed by varying the amplitude of the twisted
Zikanov mode.
A second type of perturbation is a snapshot from a turbulent run at
$\R=2150$ which is scaled in energy \ie in amplitude. A cross-section
is shown in Fig.~\ref{fig:LT_IC}. 

\begin{figure}
\begin{center}
\includegraphics[width=0.48\textwidth]{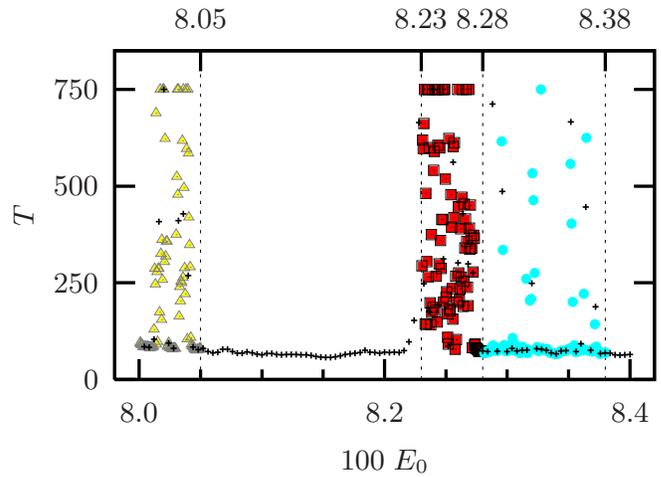} 
\end{center}
\caption[]{ (Color online) 
  Lifetime of initial conditions at $\R=1900$. The pipe is $L=15D$
  long. The initial conditions are constructed from a turbulent
  snapshot of a $\R=2150$ run (Fig.~\ref{fig:LT_IC}). It is scaled in
  kinetic energy $E_0$. The black crosses belong to an ensemble
  constructed by varying the energy form $E_0=8\times 10^{-2}$ to
  $E_0=8.4\times 10^{-2}$ in $100$ equidistant steps. 
  Additional ensembles with $100$ samples each
  focus on regions of high probability to observe long turbulent
  transients.  The yellow triangles indicate an ensemble that ranges
  form $E_0=8.0\times 10^{-2}$ to $E_0=8.05\times 10^{-2}$ in $100$
  steps.  The red squares are located in the range between
  $E_0=8.23\times 10^{-2}$ and $E_0=8.28\times 10^{-2}$.  Finally the
  cyan circles reach from $E_0=8.28\times 10^{-2}$ to $E_0=8.38\times
  10^{-2}$. The corresponding lifetime distributions for all ensembles
  are shown in Fig.~\ref{fig:LT_PDF}.
\label{fig:LT_amplitude}}
\end{figure}

Fig.~\ref{fig:LT_amplitude} shows the lifetime of the perturbation as a
function of the initial energy $E_0$. Regions of small and smoothly
varying lifetime are clearly separated from regions of longer
fluctuating lifetimes. In regions with short lifetimes the flow
relaxes quickly to the laminar profile. Towards the boundaries of
these regions the lifetimes increase quickly and reach plateaus at the
maximal integration time. Magnifications of the plateau regions show
chaotic and unpredictable variations of lifetimes \cite{Faisst2004}.
The cliff-structure in the lifetime is due to the geometric features
of the basin of attraction of the laminar profile and has been seen
in pipe flow \cite{Schneider2007a} and 
low-dimensional representations of shear flows
\cite{Moehlis2004,Moehlis2005,Skufca2006}.

A first ensemble of initial conditions is constructed by varying the
energy from $E_0=8.0\times 10^{-2}$ to $E_0=8.4\times 10^{-2}$ in $100$
equidistant steps.  Other ensembles are chosen such that they 
provide a higher resolution in initial energies for the regions where
high lifetimes are expected.

\begin{figure}
\begin{center}
\includegraphics[width=0.48\textwidth]{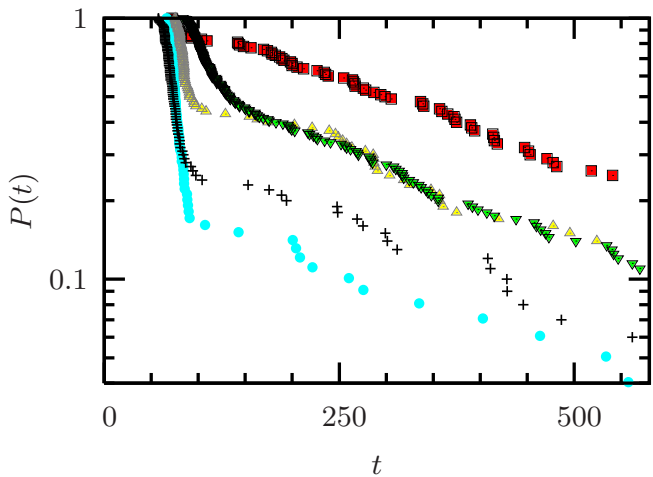} 
\end{center}
\caption[]{(Color online) 
  Lifetime distribution for fixed parameters ($\R=1900$ and periodic
  domain of length $L=15D$) for different ensembles of initial
  conditions. Presented is data obtained from an ensemble of $200$
  samples constructed from twisted Zikanov modes (green down-facing
  triangles). The other ensemble result from scaling a turbulent flow
  field from a $\R=2150$ run. The black crosses indicate an ensemble
  constructed from scaling the energy of the field from $E_0=8\times
  10^{-2}$ to $E_0=8.4\times 10^{-2}$ in $100$ equidistant steps.  The
  lifetime as a function of this energy shows typical folds (see
  Fig.~\ref{fig:LT_amplitude}). Based on this fold structure remaining
  ensembles indicated by the yellow triangles, red squares and cyan
  circles are tailored towards reaching the saddle \ie additional runs
  were chosen in the `right' energy region. The specific ensembles
  containing $100$ samples each are described in
  Fig.~\ref{fig:LT_amplitude} where the same symbols are used. While the
  initial part of the distributions depends both in length and
  functional form on the chosen ensemble, the slope of exponential
  tails is universal within the limits of statistical accuracy.
\label{fig:LT_PDF}}
\end{figure}

Fig.~\ref{fig:LT_PDF} shows $P(t)$ for fixed $\R$ calculated for the
different ensembles of initial conditions.  Each individual
distribution is characterized by an initial offset $t_0\approx 75$
where no trajectory decays. Then one observes a middle part that both
in length and functional form differs among the various ensembles
before asymptotically exponential tails are reached.  Within the
limits of statistical uncertainties the decay rates (\ie the slopes in
a semi-logarithmic plot) are independent of the ensemble of initial
conditions and encode a characteristic feature of the chaotic saddle.

In view of the cliff-structure in lifetime the origin of the
non-universal middle part becomes obvious: Only trajectories starting
from the regions of fluctuating lifetime reach the chaotic saddle from
which they can decay at a characteristic rate. Initial conditions from
the regions in between directly decay without having reached the
saddle which leads to the non-universal initial parts of $P(t)$ in
Fig.~\ref{fig:LT_PDF}. The different plots in Fig.~\ref{fig:LT_PDF} correspond
to different ensembles chosen from all possible initial conditions
presented in Fig.~\ref{fig:LT_amplitude}.

We now focus on the initial offset time $t_0$.  It corresponds to the
smallest time it takes for an initial condition from the chosen ensemble to
decay. It evidently depends on the initial perturbation. In
particular, this time scale can in principle be arbitrarily close to
zero by including an arbitrarily small perturbation to the linearly
stable laminar profile in the ensemble.
However, `typical' perturbations used both in
simulations and lab experiments 
are characterized by a
typical initial formation time $t_0$ (with $t_0\approx 100 \ldots 150$
in experiments). This can be rationalized as follows:

We first note that $t_0$ is large compared to the Lyapunov time
measured in the turbulent motion \cite{Faisst2004}, which gives a
typical timescale for the dynamical separation of neighboring
trajectories \emph{on} the saddle. However, a trajectory does not
necessarily start on the chaotic saddle. Its initial condition is a
flow field that hopefully \emph{initiates} turbulence \ie the
trajectory approaches the chaotic saddle, but typically it does not
belong to the saddle itself. In addition a trajectory starting its
decay from the saddle has to follow the evolution through state space
until it ends up in the vicinity of the laminar profile. Consequently
the offset $t_0$ contains two parts: The formation time
$t_{\text{for}}$ required to reach the turbulent `state' and the decay
time $t_{\text{dec}}$ it takes to finally reach the neighborhood of
the laminar state after the decay has been initiated. This can also be
directly observed in Fig.~\ref{fig:LT_energytraces}. The presented energy
traces consist of three parts: the initial energy growth, the chaotic
fluctuations indicating turbulent dynamics on the saddle, and the
decay towards the laminar state.


As discussed in Section \ref{sec:survey} 
the decay time can be estimated from experimental
observations that a puff decays while traveling about $50D$
  downstream \cite{Hof2004}.
This translates into a time $t_{\text{dec}}\approx 50$.
  Theoretically it follows from the mechanism of decay:
  It starts first with a reduction of transverse modulations leading
  to a break out of the regeneration cycle supporting turbulent motion
  \cite{Waleffe1997} and then shows a viscous damping of streamwise
  deviations from from the laminar profile.


The formation time $t_{\text{for}}$ can be estimated from the
experimental observation that is takes $t_{\text{for}}$ about $40$ 
for a perturbation (jet injection \cite{Hof2004}) to develop into an
equilibrium puff.  It is also observed in simulated short periodic
pipes: Taking an initial condition that contains pairs of
counter-rotating vortices in axial direction such as Zikanov's almost
optimally growing mode, these small perturbations grow in energy by
generating strong streaks in a so called lift-up process driven by the
non-normality of the Navier-Stokes operator \cite{Waleffe1995}. The
generated streaks then become unstable, transverse vortices appear and
the flow turns turbulent. The formation time can therefore be
estimated by the growing period of a Zikanov mode and originates from
the non-normal character of the Navier-Stokes operator.

The discussed mechanism is also present in experiments where the flow
is perturbed by injecting fluid jets perpendicular to the pipe
\cite{Panton2001}. These jets give rise to pairs of counter-rotating
vortices that will draw energy from the base profile and grow by the
same mechanism observed in the dynamics of a streamwise independent
Zikanov modes.  The experimental observation agrees with typical
formation times of equilibrium puffs in simulations in a periodic pipe
of length $L=50D$ where a localized form of Zikanov's mode was used as
an initial condition. 

We thus conclude that the specific choice of initial conditions
affects the functional form of the lifetime distribution for small
times. Both the initial time offset $t_0$ and the middle part of the
lifetime distribution depend on the chosen ensemble of initial
conditions.  Universal properties of the turbulent state are only
encoded in the asymptotic tails, which result from trajectories that
actually reach the turbulent state before their decay. The exponential
form of these tails is compatible with a chaotic saddle in state space
and characteristic decay rates can be `measured' by fitting
exponentials to the asymptotic tails of the lifetime distributions.

In order to probe the chaotic saddle, i.e., analyze the universal
asymptotic part of the distribution and minimize the high
computational costs at the same time, it is favorable to choose an
ensemble of initial conditions that is likely to reach turbulence.
Modulated Zikanov modes seem to be a good choice whereas independent
snapshots from a turbulent run at slightly higher $\R$ tend to decay
directly. This can for example be observed by comparing the lifetime
distributions (Fig.~\ref{fig:LT_PDF}) of the ensembles constructed by
scaling the turbulent flow field and the Zikanov mode respectively.
Although snapshots from runs at higher Reynolds number
 are complex and appear to be `turbulent'
they need not be located close to the chaotic saddle in the state
space of the system at slightly lower $\R$.  Typical length and time
scales of turbulent motion change with $\R$ so that turbulent
snapshots at one Reynolds number might not `fit' the dynamics at
another $\R$ \footnote{In simulated plane Couette flow this 
was thoroughly analyzed in 
\cite{Schmiegel2000}.  Starting form
turbulent flow at higher Reynolds number they reduced $\R$ at
different `annealing rates' and observed a direct decay when the
annealing rate was faster than intrinsic relaxation rates.}.
In contrast, in the case of Zikanov modes 
the flow has enough time to adapt to the dynamics of the
Navier-Stokes equations. Moreover, the Zikanov mode shares a
  characteristic pair of streamwise vortices with the 
\emph{edge state} of pipe flow \cite{Schneider2006,Schneider2007a}.
Since the
  edge state is located in-between the laminar state and the chaotic
  saddle, an initial condition close to the edge state should be
  especially efficient in initiating turbulence. 

  To summarize: The specific form of a lifetime distribution does not
  only depend on the system parameters such as Reynolds number and ---
  in the case of a simulation --- length of a periodic domain and
  resolution of the numerical representation that completely define
  the dynamical system.  $P(t)$ also depends on the ensemble of
  initial conditions.  The large initial offset due to transient
  growth of initial perturbations and the initial drop generated by
  trajectories that decay directly without reaching the chaotic
  `state' are not universal.  In view of the cliff-structure in
  lifetime that shifts with Reynolds-number even choosing the same
  ensemble of initial conditions does not prevent complicated
  variations of the non-universal parts of $P(t)$ with $\R$.  Only the
  exponential tails in the asymptotic regime of the lifetime
  distribution carry information about the chaotic saddle and its
  characteristic decay rate. Consequently, long observation times
  reaching into the asymptotic range and initial conditions that have
  a high probability to reach turbulent dynamics are needed in any
  study of characteristic lifetimes.

\subsection{Reynolds number dependence $\tau(\R)$}
\label{sect:variation}

Lifetime experiments were performed in periodic domains of length
$L=5D$, $L=9D$ and $L=15D$ for various Reynolds numbers. The ensemble
of initial conditions was constructed from twisted Zikanov modes of
varying amplitude. For each Reynolds number at least $100$ independent
trajectories were integrated up to a maximum integration time of $3000
R/U_{cl} = 750D/\langle{u}\rangle$ which is about twice the observation time
available in the {Manchester} pipe and a tenth of the maximal
observation time in the discussed experiments by {Hof}.
Characteristic lifetimes were extracted from the slopes of
exponentially varying tails of the measured lifetime distributions.

\begin{figure}
\begin{center}
\includegraphics[width=0.48\textwidth]{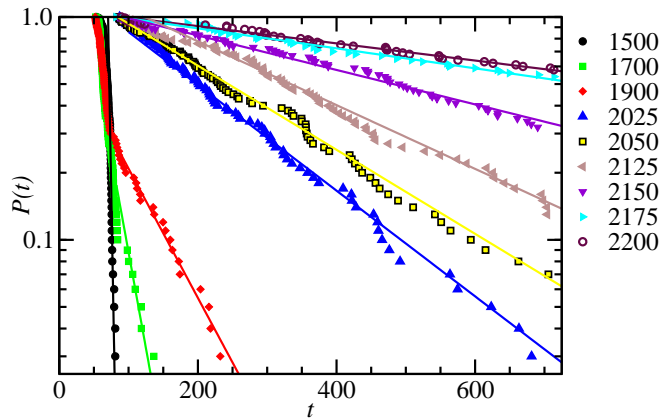} 
\end{center}
\caption[Lifetime distribution for the $L=5D$ pipe]{(Color online) 
  Probability $P(t)$ to still be turbulent after some time $t$ for the
  `short' $L=5D$ pipe. For each Reynolds number (given in the legend)
  $100$ independent initial conditions have been integrated up to a
  maximum integration time of $750$. Exponential fits to the tails of
  the distribution are indicated by straight lines in the
  semi-logarithmic representation.  The measured slopes are shown in
  Fig.~\ref{fig:LT_Tau10}.\label{fig:LT_PDF10}}
\end{figure}

\begin{figure}
\begin{center}
\includegraphics[width=0.48\textwidth]{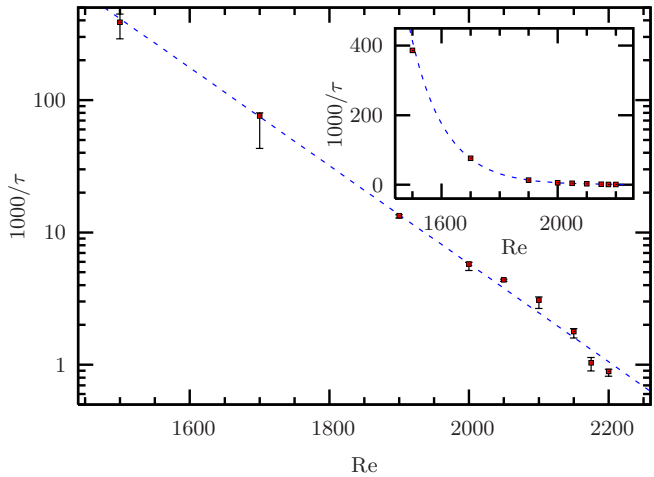} 
\end{center}
\caption[Inverse characteristic lifetimes for the $L=5D$ pipe]{(Color online) 
  Inverse characteristic lifetimes $\tau^{-1}$ for the `short' pipe of length $L=5D$,
  as extracted from the exponential fits of $P(t)$ (see
  Fig.~\ref{fig:LT_PDF10}) as a function of Reynolds number both on a
  semi-logarithmic (main) and a linear scale (inset). 
  The error bars result from extreme fits still compatible with the
  data.  Thus, they reflect both intrinsic statistical uncertainties
  of the exponential fits \emph{and} errors resulting from estimating
  the beginning of the asymptotic regime.
  The data is compatible with an exponential variation of $\tau(\R)$
  but does not support the linear scaling proposed in the literature
  before (cf. inset). Thus, the data does not support a divergence of
  $\tau$ \ie a zero crossing of $1/\tau$ at a finite Reynolds number
  close to $2000$.
  \label{fig:LT_Tau10}}
\end{figure}

Fig.~\ref{fig:LT_PDF10} shows the probability $P(t)$ to still be turbulent
at some time $t$ as a function of this time 
for the `short' $L=5 D$ pipe.  
The data points lie on
straight lines in a semi-logarithmic plot, clearly indicating an
exponential variation for large times. The slopes of the indicated fits
which correspond to $1/\tau$ with $\tau$ the characteristic timescale
of the decay process are plotted in Fig.~\ref{fig:LT_Tau10} as a function
of the Reynolds number.  

The characteristic lifetime $\tau$ increases rapidly with $\R$ which, in
previous studies \cite{Faisst2004,Mullin2006a,Peixinho2006}, 
led to the conclusion that it diverges at a finite
critical Reynolds number $\R_c$ like 
\beq
 \tau(\R) \sim \frac{1}{ \R_c - \R}\;.  
\eeq 
In a linear plot of the inverse lifetime $1/\tau$ as
a function of $\R$ this would correspond to a linear variation that
crosses zero at the critical Reynolds number. Our data do not
support this scaling. Indeed, $1/\tau$ approaches zero as we increase $\R$ but
there is no indication of a divergence. Instead the data is compatible
with an exponential scaling (\ref{exponential}),
which corresponds to a straight line in a semi-logarithmic
representation. The values for the parameters $a$ and $b$ are
listed in table~\ref{table}.

Consequently, there is no evidence for a transition from a chaotic
saddle to a permanent chaotic attractor, at least not close to a
Reynolds number of order $2000$, where critical values have been
reported previously.

Comparing with experimental results \cite{Hof2006}, our numerical studies confirm
exponential lifetime distributions and that the characteristic
lifetime does not diverge but grows exponentially with the Reynolds
number.  However, the parameters of the exponential scaling law do not
match quantitatively. This shortcoming is addressed in the next
section, where we discuss how the length of the periodic domain used
in our simulation affects the statistics.

\subsection[]{Extensitivity of $\tau(\R)$}
\label{sect:extensive}

Having found the exponential scaling of characteristic lifetimes both
in experimental works and in numerical simulation, one can
quantitatively compare both systems. The main difference between both
considered systems is that in an experiment one observes a
\emph{localized} turbulent puff traveling through a very long pipe
whereas in short simulated periodic pipes not the full spatiotemporal
structure but only the internal dynamics of a puff is captured. There
is no coexistence of a turbulent region and laminar flow and no
dynamics of the fronts of a turbulent puff. If the periodic domain
becomes long compared to all internal scales of a turbulent puff,
including its overall size, features of the experiment should be
quantitatively recovered. For shorter computational domains, however,
finite-size effects
are to be expected.

We analyze the dependence on the length $L$ of the computational
domain by comparing the results from the short $L=5D$ pipe with
additional simulations for a `medium' $L=9D$ and a `long' $L=15D$
pipe. The length $L=9D$ (rather than $L=10D$) for the medium pipe was
chosen such that periodic structures that might be favored by the
periodicity of the small reference calculation do not exactly `fit'
the new period in downstream direction.

\begin{figure}
\begin{center}
\includegraphics[width=0.48\textwidth]{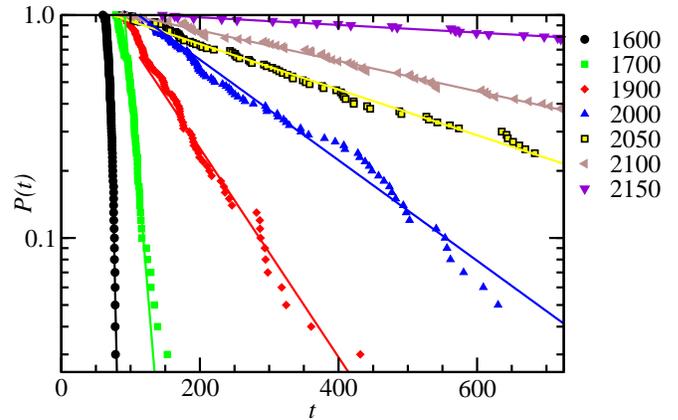} 
\end{center}
\caption[Lifetime distribution for the $L=9D$ pipe]{(Color online) 
  Lifetime distributions for the pipe of `medium' length $L=9D$ pipe. For each
  Reynolds number (legend) $100$ trajectories have been analyzed. The
  straight lines indicate exponential fits to the tails of the
  distributions.\label{fig:LT_PDF18}}
\end{figure}

Fig.~\ref{fig:LT_PDF18} shows lifetime distributions based on $100$
individual runs at every Reynolds number for the medium pipe.
Exponential fits to the tails are presented as straight lines in the
semi-logarithmic plot.
 
\begin{figure}
\begin{center}
\includegraphics[width=0.48\textwidth]{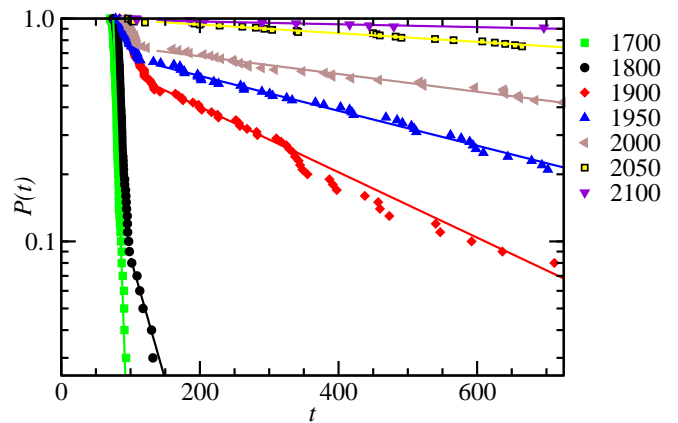} 
\end{center}
\caption[Lifetime distribution for the $L=15D$ pipe]{(Color online) 
  Same as Fig.~\ref{fig:LT_PDF18} but for the `long' pipe of length $L=15D$. 
Note the long offset $t_0$ especially for $\R=1900$. \label{fig:LT_PDF30}}
\end{figure}
Fig.~\ref{fig:LT_PDF30} shows the same data also based on the analysis of
$100$ runs at each Reynolds number for the long periodic pipe. In this
plot, the different parts of $P(t)$ \ie the initial decay
followed by asymptotic tails that have to be used for measuring the
characteristic lifetimes, are obvious.

\begin{figure}
\includegraphics[width=0.47\textwidth]{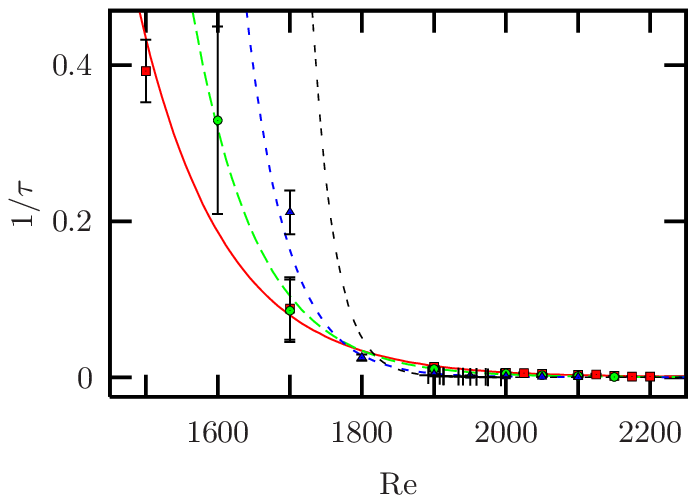} 
\qquad
\includegraphics[width=0.47\textwidth]{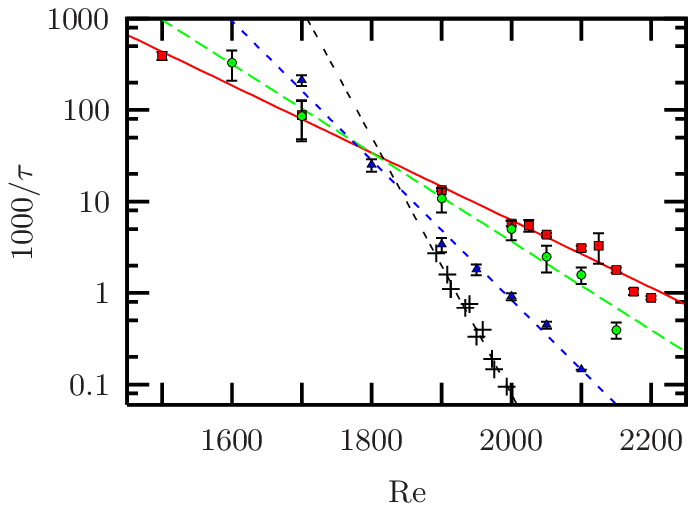} 
\caption[]{(Color online) 
  Inverse characteristic lifetimes extracted from simulations in
  periodic domains of different length ($L=5D$ (red squares), $L=9D$
  (green circles), $L=15D$ (blue triangles)) and from the experiment
  by {Hof} \cite{Hof2006} (black crosses) in a linear (left panel) and
  semi-logarithmic plot (right panel). The straight lines are
  exponential fits to the data, indicating that all data sets are
  individually compatible with an exponential scaling of lifetime with
  Reynolds number. The slope of the exponential grows with the length
  of the periodic domain and approaches quantitatively the
  experimental results.  By extrapolation shown in
  Fig.~\ref{fig:LT_lengthdep} one can speculate that numerical and
  experimental results should match at $L=30D$, which is about the
  length of an equilibrium puff.
  \label{fig:LT_Tau}}
\end{figure}

In Fig.~\ref{fig:LT_Tau} the extracted inverse characteristic lifetimes
from Fig.~\ref{fig:LT_PDF18} and Fig.~\ref{fig:LT_PDF30} are presented together
with the $L=5D$ reference data.

All three presented data sets show no evidence for a divergence. Each
one is fully compatible with an exponential scaling of $\tau$ as a
function of $\R$. However the slope of the exponential is not
universal but varies with the length of the computational domain.
Large changes in the characteristic lifetime occur in a
smaller interval of Reynolds numbers for a longer pipe. For Reynolds
numbers larger than about $1850$ the characteristic lifetime grows
with the length of the computational domain.

Hence, the characteristic lifetime is not a purely intensive measure.
It scales with the size of the system under consideration. Such a
scaling is compatible with the reasoning that has been developed
for spatially extended transient chaos, when the domains considered
are larger than the correlation length.

\begin{figure}
\begin{center}
\includegraphics[width=6.0cm,angle=0]{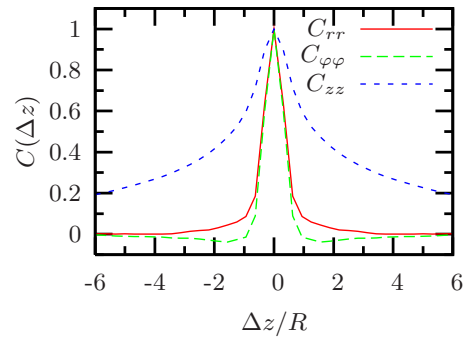}
\end{center}
\caption[]{(Color online) 
  Autocorrelation functions for the downstream ($C_{zz}$), azimuthal
  ($C_{\varphi \varphi}$) and radial ($C_{rr}$) velocities along the
  pipe axis. The correlations are evaluated in the comoving frame 
of reference of a turbulent puff \cite{statphys} and are
  normalized to one for vanishing axial shift $\Delta z$. 
\label{correlation}
}
\end{figure}

At the Reynolds numbers considered, the flow in the pipe is correlated
in azimuthal, radial and also axial direction. The axial
auto-correlation functions for the three velocity components are shown
in Fig.~\ref{correlation}.  They fall off rather quickly within about
$2R$.  The short axial correlation length suggests that useful
information about the interior dynamics of the flow can be obtained by
studying relatively short domains. Specifically, for a length of $10R$
the axial correlations in the downstream fluid are down to $30$ per
cent, but the computational advantages are enormous and allow for
detailed studies of deterministic \cite{Faisst2003} and statistical
properties \cite{Schneider2007}.

\begin{figure}
\begin{center}
\includegraphics[width=6.0cm,angle=0]{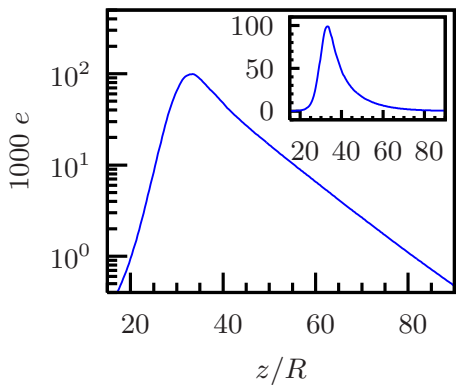}
\end{center}
\caption[]{(Color online) Energy content of a turbulent puff, averaged over time
in a co-moving frame (as discussed in \cite{statphys}) in a 
logarithmic representation (main panel) and a linear on (inset).
The rapid, exponential drop of underlines the strong localization
properties of a turbulent puff and also shows that its length
is rather well defined to be about 30 D. The spot moves 
from left to right.
\label{localization}
}
\end{figure}

\begin{figure}
\begin{center}
\includegraphics[width=0.45\textwidth]{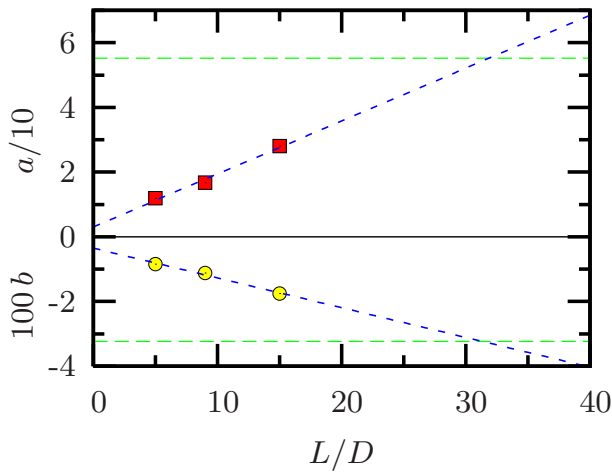} 
\end{center}
\caption[Extensitivity of characteristic lifetimes]{(Color online) 
  The variation of inverse lifetime shown in Fig.~\ref{fig:LT_Tau} may be
  described as $\tau^{-1}=\exp(a+b\;\R)$ with two parameters $a$ and
  $b$. We plot these parameters as a function of the length $L$ of the
  computational domain. The red boxes correspond to the parameter
  $a/10$ and the yellow circles to $100 b$. The experimental
  values from \cite{Hof2006} are indicated by the green dashed
  horizontal lines. Linear fits to the data points (blue dashed lines)
  reach the experimental values at approximately the same length
  $L\approx30 D$ which is about the length of an equilibrium puff.    
\label{fig:LT_lengthdep}}
\end{figure}

A turbulent puff,
on the other hand, defined via the total energy content, and averaged
over time, is much longer, see Fig.~\ref{localization}. Therefore, 
it is plausible to think of the flow in a turbulent puff as succession
of segments of length $D$, which are more or less uncorrelated. 
Let $r$ be the probability for one of them to decay over a time
interval $\delta t$. Then the probability for $N$ of them to 
decay simultaneously is $r^N$. It has to be simultaneous, for 
otherwise a turbulent cell could main turbulence and trigger 
a spreading along the axis. Since $r=1/\tau$, this implies that
the lifetime $\tau_N$ for the system of $N$ cells is
$\tau_N=\tau^N$. With the representation (\ref{exponential})
this implies a linear increase of the parameters $a$ and $b$
with the number of cells, i.e. the length of the turbulent section.
This is nicely borne out by the data in Fig.~\ref{fig:LT_lengthdep}.
 
Comparing with the characteristic lifetimes extracted from the
experiments by {Hof} included in Fig.~\ref{fig:LT_Tau} the lifetimes
extracted from simulations quantitatively approach experimental values as one
increases the length of the computational domain.

The individual exponential variation of the lifetime with $\R$ may be
described using (\ref{exponential})
with two parameters $a$ and $b$
which depend on the length $L$ of the computational domain.  In
Fig.~\ref{fig:LT_lengthdep} we plot these parameters as a function of $L$.
Linear extrapolation shows that both parameters \emph{independently}
approach the experimental values -- included as horizontal lines in
Fig.~\ref{fig:LT_lengthdep} -- at the \emph{same} length. A quantitative
match between experimental and numerical results can thus be expected
at a length of $L=30D$, which is approximately the length of
equilibrium puffs observed in long pipes.
Hence, the data suggests that the characteristic lifetime is an
extensive quantity scaling with the length of a turbulent patch. 

\begin{table}
\begin{tabular}{l||l|l}
  $L/D$ &  $a$     & $100b$ \\\hline
  $5$   &  $11.9$  & $- 0.85$ \\ 
  $9$   &  $16.7$  & $- 1.11$ \\ 
 $15$   &  $28.0$  & $- 1.76$ \\\hline 
30 (extrap.) & $52.1$ & $- 3.11$ \\\hline 
experiment  & $55.2$  & $-3.23$\\ 
\end{tabular}
\caption[]{Parameters for the exponential law (\ref{exponential}).
The top three rows are from numerical simulations of periodic sections
of length $L=5D$, $9D$ and $15D$, respectively. The next to last line
gives the value extrapolated for $L=30D$ and the last row gives the
experimental data of \cite{Hof2004}.}
\label{table}
\end{table}

Our findings support the assumption that turbulent puffs do not decay
from their boundaries but decay is initiated in the bulk region. Short
periodic pipes appear to faithfully reproduce the internal dynamics of
a turbulent puff and the dynamics of the fronts is of minor importance.

\section{Conclusions}
\label{sect:end}

The results presented here confirm that lifetime distributions 
of turbulent pipe flow asymptotically follow exponentials. 
While there may be differences for short times, the long time behavior is robust. 
This observation supports the idea that
turbulent motion is generated by a chaotic saddle in state space.
Features of the measured probability functions can be explained in
terms of an ensemble of initial conditions that either directly decay
or reach the strange chaotic saddle. From this saddle, 
trajectories decay at a constant escape rate, i.e., independent of the 
previous state. This situation is analogous to that of a particle 
moving in a complex box with some holes through which it can decay
\cite{Ott2002,Faisst2004},
or an unstable nucleus subject to radioactive decay. After escaping
from the saddle trajectories then follow the slow dynamics towards the
linearly stable laminar profile.

A thorough analysis of lifetime distributions from both experimental
and numerical studies of pipe flow as well as plane Couette flow
shows that the characteristic lifetimes grow with Reynolds number, but
that they do not diverge at a finite value of Reynolds numbers. As a 
consequence, there is no evidence that the chaotic saddle
in state space turns into an attractor by some sort of `inverse
boundary crisis' \cite{Ott2002}. 
Even for Reynolds numbers exceeding $2000$ turbulent
signals consist of transients that decay finally, though the time for 
decay can be very long. As regards much higher Reynolds numbers and in
particular the transition from puffs to slugs which increase in length,
the present results suggest that also the decay time $\tau(Re)$ may become longer.
But this is insufficient to suggest a transition to a permanent attractor, so
that the question about a global bifurcation that turns the turbulent dynamics
into an attractor remains open.
In any case, the fact that turbulent motion stays dynamically connected
with the laminar profile at Reynolds numbers exceeding $2000$ suggests
that turbulent flow could be intentionally laminarized at minimal
energetic costs. 

An intriguing finding of the present study is a dependence of the 
characteristic lifetimes on the spatial extension of a
turbulent region. The larger it is the less likely it decays.  
For the case of puffs, it is possible to extrapolate the present numerical
results to the experimental ones. The results for pipe flow are consistent
with theoretical models for spatially extended systems. It will be interesting
to see to which extend the extensive scaling of lifetimes seen here in pipe flow
is generic and can be found in other linearly stable flows such as plane Couette flow.

\section*{Acknowledgments}
We thank Bjorn Hof and Jerry Westerweel for stimulating discussions.
This work was supported in part by Deutsche Forschungsgemeinschaft.




\end{document}